\documentclass[italian,english]{article}
\usepackage[T1]{fontenc}
\usepackage[latin1]{inputenc}
\usepackage{color}
\usepackage{graphicx}
\usepackage{amssymb}

\makeatletter

 \newcommand{\lyxaddress}[1]{
   \par {\raggedright #1 
   \vspace{1.4em}
   \noindent\par}
 }

\usepackage{babel}
\makeatother
\begin{document}

\title{\textbf{The production of matter from curvature in a particular linearized
high order theory of gravity and the longitudinal response function
of interferometers }}

\author{\textbf{Christian Corda}}

\maketitle

\lyxaddress{\begin{center}INFN - Sezione di Pisa and Università di Pisa, Via
F. Buonarroti 2, I - 56127 PISA, Italy\end{center}}

\lyxaddress{\begin{center}\textit{E-mail address:} \textcolor{blue}{christian.corda@ego-gw.it} \end{center}}

\begin{abstract}
The strict analogy between scalar-tensor theories of gravity and high
order gravity is well known in literature. In this paper it is shown
that, from a particular high order gravity theory known in literature,
it is possible to produce, in the linearized approch, particles which
can be seen like massive scalar modes of gravitational waves and the
response of interferometers to this type of particles is analyzed.
The presence of the mass generates a longitudinal force in addition
of the transverse one which is proper of the massless gravitational
waves and the response of an arm of an interferometer to this longitudinal
effect in the frame of a local observer is computed. This longitudinal
response function is directly connected with the function of the Ricci
scalar in the particular action of this high order theory. Important
conseguences from a theoretical point of view could arise from this
approach, because it opens to the possibility of using the signals
seen from interferometers to understand which is the correct theory
of gravitation.

The presence of the mass could also have important applications in
cosmology because the fact that gravitational waves can have mass
could give a contribution to the dark matter of the Universe.
\end{abstract}

\lyxaddress{PACS numbers: 04.80.Nn, 04.30.Nk, 04.50.+h}

\section{Introduction}

The design and construction of a number of sensitive detectors for
gravitational waves is underway today. There are some laser interferometers
like the Virgo detector, being built in Cascina, near Pisa, Italy,
by a joint Italian-French collaboration, the GEO 600 detector being
built in Hannover, Germany, by a joint Anglo-Germany collaboration,
the two LIGO detectors being built in the United States (one in Hanford,
Washington and the other in Livingston, Louisiana) by a joint Caltech-Mit
collaboration, and the TAMA 300 detector, being built near Tokyo,
Japan. Many bar detectors are currently in operation too, and several
interferometers and bars are in a phase of planning and proposal stages
(for the current status of gravitational waves experiments see \cite{key-1,key-2}).

The results of these detectors will have a fundamental impact on astrophysics
and gravitation physics. There will be lots of experimental data to
be analyzed, and theorists will be forced to interact with lots of
experiments and data analysts to extract the physics from the data
stream.

In this paper the production and the potential detection with interferometers
of a hypotetical massive \textit{scalar} component of gravitational
radiation which arises from a particular high order theory of gravity
well known in literature (see \cite{key-3}) is shown. This agrees
with the formal equivalence between high order theories of gravity
and scalar tensor gravity which is well known in literature \cite{key-3,key-4,key-5,key-6}. 

In the second Section of this paper it is shown that a massive scalar
mode of gravitational radiation arises from the high order action
\cite{key-3}\begin{equation}
S=\int d^{4}x\sqrt{-g}R^{-1}+\mathcal{L}_{m},\label{eq: high order 1}\end{equation}

where $R$ is the Ricci scalar curvature. Equation (\ref{eq: high order 1})
is a particular choice with respect the well known canonical one of
General Relativity (the Einstein - Hilbert action \cite{key-7,key-8})
which is 

\begin{equation}
S=\int d^{4}x\sqrt{-g}R+\mathcal{L}_{m}.\label{eq: EH}\end{equation}
The presence of the mass could also have important applications in
cosmology because the fact that gravitational waves can have mass
could give a contribution to the dark matter of the Universe. We also
recall that an alternative way to resolve the dark matter and dark
energy problems using high order gravity is shown in ref. \cite{key-9}.

In Section three it is shown that this massive component generates
a longitudinal force in addition of the transverse one which is proper
of the massless case. 

After this, in Section four, the potential interferometric detection
of this longitudinal component is analyzed and the response of an
interferometer is computed. It is also shown that this longitudinal
response function is directly connected with the Ricci scalar $R$.
This connection opens to the possibility of using the signals seen
from interferometers to understand which is the correct theory of
gravitation \cite{key-6,key-10}.

In the analysis of the longitudinal response of interferometers a
computation parallel to the one seen in \cite{key-6} and \cite{key-11}
will be used.

\section{The production of a scalar massive mode of gravitational radiation
in the $R^{-1}$ high order theory of gravity}

If the gravitational Lagrangian is nonlinear in the curvature invariants
the Einstein field equations has an order higher than second \cite{key-4,key-6}.
For this reason such theories are often called higher-order gravitational
theories. This is exactly the case of the action (\ref{eq: high order 1}).

By varying this action with respect to $g_{\mu\nu}$ (see refs. \cite{key-4,key-6}
for a parallel computation) the field equations are obtained (note
in this paper we work with $G=1$, $c=1$ and $\hbar=1$):

\begin{equation}
\begin{array}{c}
G_{\mu\nu}=4\pi\tilde{G}R^{2}T_{\mu\nu}^{(m)}-\frac{R^{5}}{4}((R^{-2})_{;\mu}(R^{-2})_{;\nu}-\frac{1}{2}g_{\mu\nu}g^{\alpha\beta}(R^{-2})_{;\alpha}(R^{-2})_{;\beta})+\\
\\+R^{2}((R^{-2})_{;\mu\nu}-g_{\mu\nu}[]R^{-2})-\frac{R^{2}}{2}g_{\mu\nu}g^{\mu\nu}(R^{-2})_{;\mu}(R^{-2})_{;\nu}\end{array}\label{eq: einstein-general}\end{equation}

with associed a Klein - Gordon equation for the $R^{-2}$ scalar field

\begin{equation}
\begin{array}{c}
[]R^{-2}=\frac{-2}{-R+6}[-4\pi\tilde{G}T^{(m)}+g^{\mu\nu}(R^{-2})_{;\mu}(R^{-2})_{;\nu}+\\
\\-R^{-2}\frac{d}{d(R^{-2})}(\frac{1}{2}g^{\mu\nu}(R^{-2})_{;\mu}(R^{-2})_{;\nu})+\frac{1}{4}\frac{dR}{d(R^{-2})}g^{\mu\nu}(R^{-2})_{;\mu}(R^{-2})_{;\nu}],\end{array}\label{eq: KG}\end{equation}

where $[]$ is the Dalembertian operator.

In the above equations $T_{\mu\nu}^{(m)}$ is the ordinary stress-energy
tensor of the matter and $\tilde{G}$ is a dimensional, strictly positive,
constant \cite{key-4,key-6}. The Newton constant is replaced by the
effective coupling

\begin{equation}
G_{eff}=-\frac{R^{2}}{2},\label{eq: newton eff}\end{equation}

which is different from $G$. 

To study gravitational waves the linearized theory in vacuum ($T_{\mu\nu}^{(m)}=0$)
has to be analyzed, with a little perturbation of the background,
which is assumed given by the Minkowskian background plus $R^{-2}=(R^{-2})_{0}$
(i.e., the Ricci scalar is assumed constant). 

In the linearized theory it is also

\begin{equation}
\frac{1}{2}g^{\mu\nu}(R^{-2})_{;\mu}(R^{-2})_{;\nu}\simeq\frac{1}{2}\eta^{\mu\nu}(R^{-2})_{;\mu}(R^{-2})_{;\nu}\label{eq: V(R) 2}\end{equation}

where $\eta^{\mu\nu}$ is the flat metric tensor of the Minkowskian
background. Thus $\frac{1}{2}g^{\mu\nu}(R^{-2})_{;\mu}(R^{-2})_{;\nu}$
is proportional to the square of the total covariant derivative in
good approximation. It can also be written like

\begin{equation}
\begin{array}{c}
\frac{1}{2}g^{\mu\nu}(R^{-2})_{;\mu}(R^{-2})_{;\nu}\simeq\beta\delta(-R^{-2})^{2}\Rightarrow\\
\\\Rightarrow\frac{d}{d(R^{-2})}(\frac{1}{2}g^{\mu\nu}(R^{-2})_{;\mu}(R^{-2})_{;\nu})\simeq2\beta\delta(-R^{-2}).\end{array}\label{eq: minimo 2}\end{equation}

Putting

\begin{equation}
\begin{array}{c}
g_{\mu\nu}=\eta_{\mu\nu}+h_{\mu\nu}\\
\\(-R^{-2})_{*}=-(R^{-2})_{0}+\delta(-R^{-2}),\end{array}\label{eq: linearizza}\end{equation}

to first order in $h_{\mu\nu}$ and $\delta(-R^{-2})$, calling $\widetilde{R}_{\mu\nu\rho\sigma}$
, $\widetilde{R}_{\mu\nu}$ and $\widetilde{R}$ the linearized quantity
which correspond to $R_{\mu\nu\rho\sigma}$ , $R_{\mu\nu}$ and $R$,
the linearized field equations are obtained \cite{key-6,key-8}:

\begin{equation}
\begin{array}{c}
\widetilde{R}_{\mu\nu}-\frac{\widetilde{R}}{2}\eta_{\mu\nu}=\partial_{\mu}\partial_{\nu}\xi-\eta_{\mu\nu}[]\xi\\
\\{}[]\xi=-m^{2}\xi,\end{array}\label{eq: linearizzate1}\end{equation}

where 

\begin{equation}
\begin{array}{c}
\xi\equiv\frac{\delta(-R^{-2})}{(-R^{-2})_{0}}\\
\\m^{2}\equiv\frac{-\beta(\frac{R}{4})_{0}}{-2(\frac{R}{4})_{0}+3}=\frac{\beta R_{0}}{2R_{0}-12}.\end{array}\label{eq: definizione}\end{equation}

We emphasize that the mass is directly generated by a function of
the Ricci scalar (i.e. by curvature).

$\widetilde{R}_{\mu\nu\rho\sigma}$ and eqs. (\ref{eq: linearizzate1})
are invariants for gauge transformations \cite{key-6,key-8}

\begin{equation}
\begin{array}{c}
h_{\mu\nu}\rightarrow h'_{\mu\nu}=h_{\mu\nu}-\partial_{(\mu}\epsilon_{\nu)}\\
\\\delta(R^{-2})\rightarrow\delta(R^{-2})'=\delta(R^{-2});\end{array}\label{eq: gauge}\end{equation}

then 

\begin{equation}
\bar{h}_{\mu\nu}\equiv h_{\mu\nu}-\frac{h}{2}\eta_{\mu\nu}-\eta_{\mu\nu}\xi\label{eq: ridefiniz}\end{equation}

can be defined, and, considering the transform for the parameter $\epsilon^{\mu}$

\begin{equation}
[]\epsilon_{\nu}=\partial^{\mu}\bar{h}_{\mu\nu},\label{eq:lorentziana}\end{equation}
 a gauge parallel to the Lorenz one of electromagnetic waves can be
choosen:

\begin{equation}
\partial^{\mu}\bar{h}_{\mu\nu}=0.\label{eq: cond lorentz}\end{equation}

In this way field equations read like

\begin{equation}
[]\bar{h}_{\mu\nu}=0\label{eq: onda T}\end{equation}

\begin{equation}
[]\xi=-m^{2}\xi.\label{eq: onda S}\end{equation}

Solutions of eqs. (\ref{eq: onda T}) and (\ref{eq: onda S}) are
plan waves:

\begin{equation}
\bar{h}_{\mu\nu}=A_{\mu\nu}(\overrightarrow{p})\exp(ip^{\alpha}x_{\alpha})+c.c.\label{eq: sol T}\end{equation}

\begin{equation}
\xi=a(\overrightarrow{p})\exp(iq^{\alpha}x_{\alpha})+c.c.\label{eq: sol S}\end{equation}

where

\begin{equation}
\begin{array}{ccc}
k^{\alpha}\equiv(\omega,\overrightarrow{p}) &  & \omega=p\equiv|\overrightarrow{p}|\\
\\q^{\alpha}\equiv(\omega_{m},\overrightarrow{p}) &  & \omega_{m}=\sqrt{m^{2}+p^{2}}.\end{array}\label{eq: k e q}\end{equation}

In eqs. (\ref{eq: onda T}) and (\ref{eq: sol T}) the equation and
the solution for the tensorial waves exactly like in General Relativity
\cite{key-7,key-8} have been obteined, while eqs. (\ref{eq: onda S})
and (\ref{eq: sol S}) are respectively the equation and the solution
for the scalar mode (see also \cite{key-6}).

The fact that the dispersion law for the modes of the scalar massive
field $\xi$ is not linear has to be emphatized. The velocity of every
tensorial mode $\bar{h}_{\mu\nu}$ is the light speed $c$, but the
dispersion law (the second of eq. (\ref{eq: k e q})) for the modes
of $\xi$ is that of a massive field which can be discussed like a
wave-packet \cite{key-6,key-12}. Also, the group-velocity of a wave-packet
of $\xi$ centered in $\overrightarrow{p}$ is 

\begin{equation}
\overrightarrow{v_{G}}=\frac{\overrightarrow{p}}{\omega},\label{eq: velocita' di gruppo}\end{equation}

which is exactly the velocity of a massive particle with mass $m$
and momentum $\overrightarrow{p}$.

From the second of eqs. (\ref{eq: k e q}) and eq. (\ref{eq: velocita' di gruppo})
it is simple to obtain:

\begin{equation}
v_{G}=\frac{\sqrt{\omega^{2}-m^{2}}}{\omega}.\label{eq: velocita' di gruppo 2}\end{equation}

Then, wanting a constant speed of our wave-packet, it has to be \cite{key-6}

\begin{equation}
m=\sqrt{(1-v_{G}^{2})}\omega.\label{eq: relazione massa-frequenza}\end{equation}

The relation (\ref{eq: relazione massa-frequenza}) is shown in fig.
1 for a value $v_{G}=0.9$.

\begin{figure}
\includegraphics{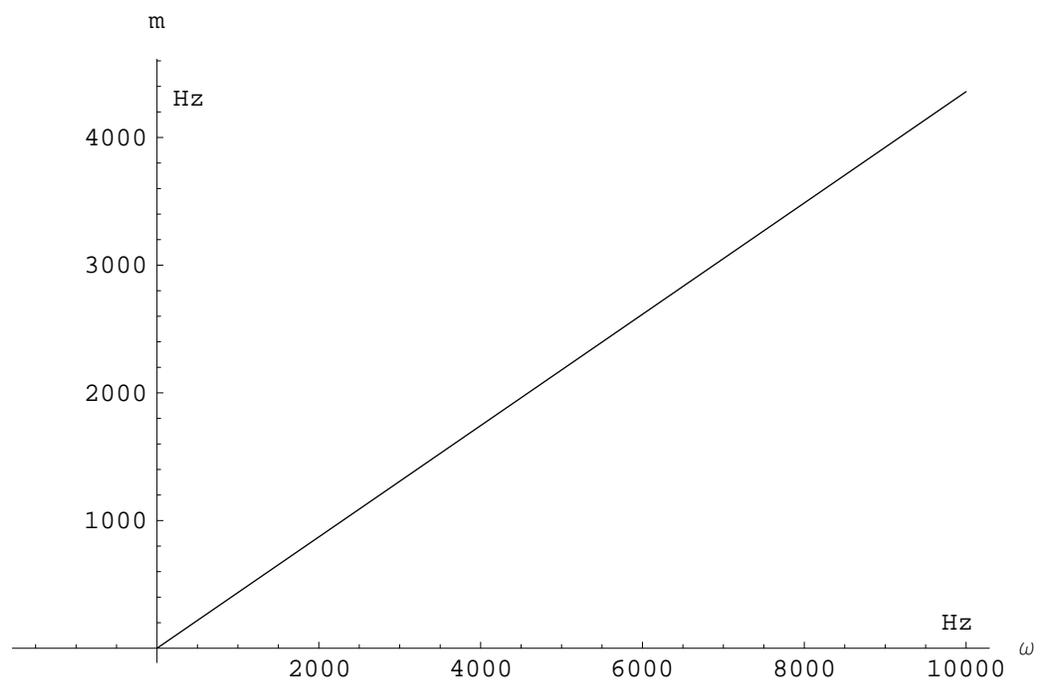}

\caption{the mass-frequency relation for a massive SGW propagating with a
speed of $0.9c$ : for the mass it is $1Hz=10^{-15}eV$}
\end{figure}

Now the analisys can remain in the Lorenz gauge with trasformations
of the type $[]\epsilon_{\nu}=0$; this gauge gives a condition of
transversality for the tensorial part of the field: $k^{\mu}A_{\mu\nu}=0$,
but does not give the transversality for the total field $h_{\mu\nu}$.
From eq. (\ref{eq: ridefiniz}) it is

\begin{equation}
h_{\mu\nu}=\bar{h}_{\mu\nu}-\frac{\bar{h}}{2}\eta_{\mu\nu}-\eta_{\mu\nu}\xi.\label{eq: ridefiniz 2}\end{equation}

At this point, if being in the massless case \cite{key-6}, it could
been put

\begin{equation}
\begin{array}{c}
[]\epsilon^{\mu}=0\\
\\\partial_{\mu}\epsilon^{\mu}=-\frac{\bar{h}}{2}-\xi,\end{array}\label{eq: gauge2}\end{equation}

which gives the total transversality of the field. But in the massive
case this is impossible. In fact, applying the Dalembertian operator
to the second of eqs. (\ref{eq: gauge2}) and using the field equations
(\ref{eq: onda T}) and (\ref{eq: onda S}) it results

\begin{equation}
[]\epsilon^{\mu}=-m^{2}\xi,\label{eq: contrasto}\end{equation}

which is in contrast with the first of eqs. (\ref{eq: gauge2}). In
the same way it is possible to show that it does not exist any linear
relation between the tensorial field $\bar{h}_{\mu\nu}$ and the scalar
field $\xi$. Thus a gauge in wich $h_{\mu\nu}$ is purely spatial
cannot be chosen (i.e. it cannot be put $h_{\mu0}=0,$ see eq. (\ref{eq: ridefiniz 2}))
. But the traceless condition to the field $\bar{h}_{\mu\nu}$ can
be put :

\begin{equation}
\begin{array}{c}
[]\epsilon^{\mu}=0\\
\\\partial_{\mu}\epsilon^{\mu}=-\frac{\bar{h}}{2}.\end{array}\label{eq: gauge traceless}\end{equation}

These equations imply

\begin{equation}
\partial^{\mu}\bar{h}_{\mu\nu}=0.\label{eq: vincolo}\end{equation}

To save the conditions $\partial_{\mu}\bar{h}^{\mu\nu}$ and $\bar{h}=0$
transformations like

\begin{equation}
\begin{array}{c}
[]\epsilon^{\mu}=0\\
\\\partial_{\mu}\epsilon^{\mu}=0\end{array}\label{eq: gauge 3}\end{equation}

can be used and, taking $\overrightarrow{p}$ in the $z$ direction,
a gauge in which only $A_{11}$, $A_{22}$, and $A_{12}=A_{21}$ are
different to zero can be chosen. The condition $\bar{h}=0$ gives
$A_{11}=-A_{22}$. Now, putting these equations in eq. (\ref{eq: ridefiniz 2})
and defining $\Phi\equiv-\xi$ it results

\begin{equation}
h_{\mu\nu}(t,z)=A^{+}(t-z)e_{\mu\nu}^{(+)}+A^{\times}(t-z)e_{\mu\nu}^{(\times)}+\Phi(t-v_{G}z)\eta_{\mu\nu}.\label{eq: perturbazione totale}\end{equation}

The term $A^{+}(t-z)e_{\mu\nu}^{(+)}+A^{\times}(t-z)e_{\mu\nu}^{(\times)}$
describes the two standard (i.e. tensorial) polarizations of gravitational
waves which arise from General Relativity, while the term $\Phi(t-v_{G}z)\eta_{\mu\nu}$
is the scalar massive field arising from the high order theory.

\section{The origin of a longitudinal force}

For a purely scalar gravitational wave eq. (\ref{eq: perturbazione totale})
can be rewritten as

\begin{equation}
h_{\mu\nu}(t-v_{G}z)=\Phi(t-v_{G}z)\eta_{\mu\nu}\label{eq: perturbazione scalare}\end{equation}
and the corrispondent line element is

\begin{equation}
ds^{2}=[1+\Phi(t-v_{G}z)](-dt^{2}+dz^{2}+dx^{2}+dy^{2}).\label{eq: metrica puramente scalare}\end{equation}
But, in a laboratory environment on Earth, the coordinate system in
which the space-time is locally flat \cite{key-6,key-7,key-8} is
typically used and the distance between any two points is given simply
by the difference in their coordinates in the sense of Newtonian physics.
This frame is the proper reference frame of a local observer, located
for example in the position of the beam splitter of an interferometer.
In this frame gravitational waves manifest themself by exerting tidal
forces on the masses (the mirror and the beam-splitter in the case
of an interferometer). A detailed analysis of the frame of the local
observer is given in ref. \cite{key-8}, sect. 13.6. Here only the
more important features of this coordinate system are remembered:

the time coordinate $x_{0}$ is the proper time of the observer O;

spatial axes are centered in O;

in the special case of zero acceleration and zero rotation the spatial
coordinates $x_{j}$ are the proper distances along the axes and the
frame of the local observer reduces to a local Lorentz frame: in this
case the line element reads \cite{key-8}

\begin{equation}
ds^{2}=-(dx^{0})^{2}+\delta_{ij}dx^{i}dx^{j}+O(|x^{j}|^{2})dx^{\alpha}dx^{\beta}.\label{eq: metrica local lorentz}\end{equation}

The effect of the gravitational wave on test masses is described by
the equation

\begin{equation}
\ddot{x^{i}}=-\widetilde{R}_{0k0}^{i}x^{k},\label{eq: deviazione geodetiche}\end{equation}
which is the equation for geodesic deviation in this frame.

Thus, to study the effect of the scalar gravitational wave on test
masses, $\widetilde{R}_{0k0}^{i}$ has to be computed in the proper
reference frame of the local observer. But, because the linearized
Riemann tensor $\widetilde{R}_{\mu\nu\rho\sigma}$ is invariant under
gauge transformations \cite{key-6,key-8,key-12}, it can be directly
computed from eq. (\ref{eq: perturbazione scalare}). 

From \cite{key-8} it is:

\begin{equation}
\widetilde{R}_{\mu\nu\rho\sigma}=\frac{1}{2}\{\partial_{\mu}\partial_{\beta}h_{\alpha\nu}+\partial_{\nu}\partial_{\alpha}h_{\mu\beta}-\partial_{\alpha}\partial_{\beta}h_{\mu\nu}-\partial_{\mu}\partial_{\nu}h_{\alpha\beta}\},\label{eq: riemann lineare}\end{equation}

that, in the case eq. (\ref{eq: perturbazione scalare}), begins

\begin{equation}
\widetilde{R}_{0\gamma0}^{\alpha}=\frac{1}{2}\{\partial^{\alpha}\partial_{0}\Phi\eta_{0\gamma}+\partial_{0}\partial_{\gamma}\Phi\delta_{0}^{\alpha}-\partial^{\alpha}\partial_{\gamma}\Phi\eta_{00}-\partial_{0}\partial_{0}\Phi\delta_{\gamma}^{\alpha}\};\label{eq: riemann lin scalare}\end{equation}

the different elements are (only the non zero ones will be written):

\begin{equation}
\partial^{\alpha}\partial_{0}\Phi\eta_{0\gamma}=\left\{ \begin{array}{ccc}
\partial_{t}^{2}\Phi & for & \alpha=\gamma=0\\
\\-\partial_{z}\partial_{t}\Phi & for & \alpha=3;\gamma=0\end{array}\right\} \label{eq: calcoli}\end{equation}

\begin{equation}
\partial_{0}\partial_{\gamma}\Phi\delta_{0}^{\alpha}=\left\{ \begin{array}{ccc}
\partial_{t}^{2}\Phi & for & \alpha=\gamma=0\\
\\\partial_{t}\partial_{z}\Phi & for & \alpha=0;\gamma=3\end{array}\right\} \label{eq: calcoli2}\end{equation}

\begin{equation}
-\partial^{\alpha}\partial_{\gamma}\Phi\eta_{00}=\partial^{\alpha}\partial_{\gamma}\Phi=\left\{ \begin{array}{ccc}
-\partial_{t}^{2}\Phi & for & \alpha=\gamma=0\\
\\\partial_{z}^{2}\Phi & for & \alpha=\gamma=3\\
\\-\partial_{t}\partial_{z}\Phi & for & \alpha=0;\gamma=3\\
\\\partial_{z}\partial_{t}\Phi & for & \alpha=3;\gamma=0\end{array}\right\} \label{eq: calcoli3}\end{equation}

\begin{equation}
-\partial_{0}\partial_{0}\Phi\delta_{\gamma}^{\alpha}=\begin{array}{ccc}
-\partial_{z}^{2}\Phi & for & \alpha=\gamma\end{array}.\label{eq: calcoli4}\end{equation}

Now, putting these results in eq. (\ref{eq: riemann lin scalare})
it results:

\begin{equation}
\begin{array}{c}
\widetilde{R}_{010}^{1}=-\frac{1}{2}\ddot{\Phi}\\
\\\widetilde{R}_{010}^{2}=-\frac{1}{2}\ddot{\Phi}\\
\\\widetilde{R}_{030}^{3}=\frac{1}{2}[]\Phi.\end{array}\label{eq: componenti riemann}\end{equation}

But, putting the field equation (\ref{eq: onda S}) in the third of
eqs. (\ref{eq: componenti riemann}) it is

\begin{equation}
\widetilde{R}_{030}^{3}=\frac{1}{2}m^{2}\Phi,\label{eq: terza riemann}\end{equation}

which shows that the field is not transversal. 

Infact, using eq. (\ref{eq: deviazione geodetiche}) it results

\begin{equation}
\ddot{x}=\frac{1}{2}\ddot{\Phi}x,\label{eq: accelerazione mareale lungo x}\end{equation}

\begin{equation}
\ddot{y}=\frac{1}{2}\ddot{\Phi}y\label{eq: accelerazione mareale lungo y}\end{equation}

and 

\begin{equation}
\ddot{z}=-\frac{1}{2}m^{2}\Phi(t-v_{G}z)z.\label{eq: accelerazione mareale lungo z}\end{equation}

Then the effect of the mass is the generation of a \textit{longitudinal}
force (in addition to the transverse one). Note that in the limit
$m\rightarrow0$ the longitudinal force vanishes.

\section{Analysis of the interferometer's response to the longitudinal component}

Before starting the analysis it has to be discussed if there are fenomenogical
limitations to the mass of the scalar particle \cite{key-12,key-13}.
Treating scalars like classical waves, that act coherently with the
interferometer, it has to be $m\ll1/L$ , where $L=3$ kilometers
in the case of Virgo and $L=4$ kilometers in the case of LIGO \cite{key-1,key-2,key-6}.
Thus it has to be approximately $m<10^{-9}eV$. However there is a
stronger limitation coming from the fact that the scalar wave needs
a frequency which falls in the frequency-range for earth based gravitational
antennas that is the interval $10Hz\leq f\leq10KHz$ \cite{key-1,key-2}.
For a massive scalar gravitational wave, from the second of eqs. (\ref{eq: k e q})
it is:

\begin{equation}
2\pi f=\omega=\sqrt{m^{2}+p^{2}},\label{eq: frequenza-massa}\end{equation}

were $p$ is the momentum \cite{key-13}. Thus it needs

\begin{equation}
0eV\leq m\leq10^{-11}eV.\label{eq: range di massa}\end{equation}

For these light scalars their effect can be still discussed as a coherent
gravitational wave.

Equations (\ref{eq: accelerazione mareale lungo x}), (\ref{eq: accelerazione mareale lungo y})
and (\ref{eq: accelerazione mareale lungo z}) give the tidal acceleration
of the test mass caused by the scalar gravitational wave respectly
in the $x$ direction, in the $y$ direction and in the $z$ direction
\cite{key-6,key-11}.

Equivalently we can say that there is a gravitational potential \cite{key-6,key-8,key-11}:

\begin{equation}
V(\overrightarrow{r},t)=-\frac{1}{4}\ddot{\Phi}(t-\frac{z}{v_{P}})[x^{2}+y^{2}]+\frac{1}{2}m^{2}\int_{0}^{z}\Phi(t-v_{G}z)ada,\label{eq:potenziale in gauge Lorentziana generalizzato}\end{equation}

which generates the tidal forces, and that the motion of the test
mass is governed by the Newtonian equation

\begin{equation}
\ddot{\overrightarrow{r}}=-\bigtriangledown V.\label{eq: Newtoniana}\end{equation}

To obtain the longitudinal component of the scalar gravitational wave
the solution of eq. (\ref{eq: accelerazione mareale lungo z}) has
to be found. 

For this goal the perturbation method can be used. A function of time
for a fixed $z$, $\psi(t-v_{G}z)$,  can be defined \cite{key-6},
for which it is

\begin{equation}
\ddot{\psi}(t-v_{G}z)\equiv\Phi(t-v_{G}z)\label{eq: definizione di psi}\end{equation}

(note: the most general definition is $\psi(t-v_{G}z)+a(t-v_{G}z)+b$,
but, assuming only small variatons in the positions of the test masses,
it results $a=b=0$).

In this way it results

\begin{equation}
\delta z(t-v_{G}z)=-\frac{1}{2}m^{2}z_{0}\psi((t-v_{G}z).\label{eq: spostamento lungo z}\end{equation}

A feature of the frame of a local observer is the coordinate dependence
of the tidal forces due by scalar gravitational waves which can be
changed with a mere shift of the origin of the coordinate system \cite{key-6,key-11}:

\begin{equation}
x\rightarrow x+x',\textrm{ }y\rightarrow y+y'\textrm{ and }z\rightarrow z+z'.\label{eq: shift coordinate}\end{equation}

The same applies to the test mass displacements, in the $z$ direction,
eq. (\ref{eq: spostamento lungo z}). This is an indication that the
coordinates of a local observer are not simple as they could seem
\cite{key-6,key-8,key-11}. 

Now, let us consider the relative motion of test masses. A good way
to analyze variations in the proper distance (time) of test masses
is by means of {}``bouncing photons'' (see refs. \cite{key-6,key-11}
and figure 2). %
\begin{figure}
\includegraphics{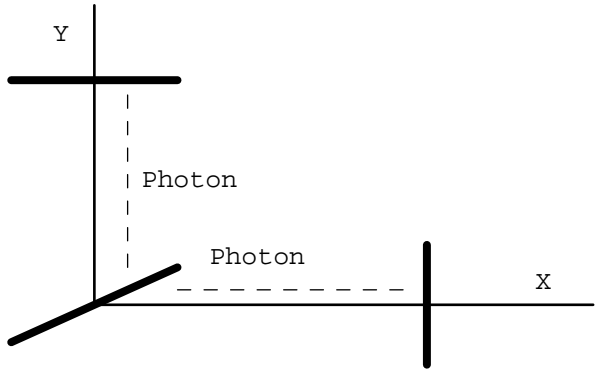}

\caption{photons can be launched from the beam-splitter to be bounced back
by the mirror}
\end{figure}
A photon can be launched from the beam-splitter to be bounced back
by the mirror. It will be assumed that both the beam-splitter and
the mirror are located along the $z$ axis of our coordinate system
(i.e. an arm of the interferometer is in the $z$ direction, which
is the direction of the propagating massive scalar gravitational wave
and of the longitudinal force, see also Figure 3).

\begin{figure}
\includegraphics{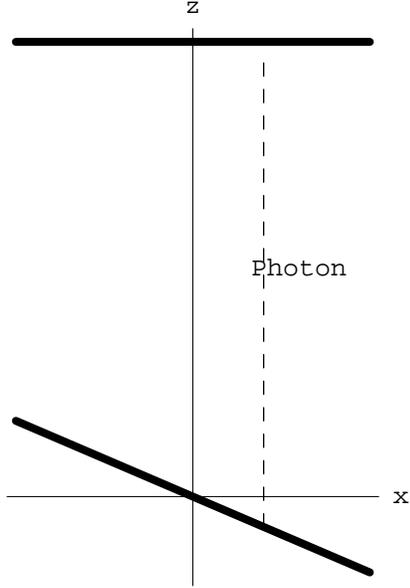}

\caption{the beam splitter and the mirror are located in the direction of
the incoming SGW}
\end{figure}

It will be shown that, in the frame of a local observer, two different
effects have to be considered in the calculation of the variation
of the round-trip time for photons, in analogy with the cases of \cite{key-6}
and \cite{key-11}, where the considered effects were three, but,
if we put the beam splitter in the origin of our coordinate system,
the third effect vanishes. 

The unperturbed coordinates for the beam-splitter and the mirror are
$x_{b}=0$ and $x_{m}=L$. So the unperturbed propagation time between
the two masses is

\begin{equation}
T=L.\label{eq: tempo imperturbato}\end{equation}

From eq. (\ref{eq: spostamento lungo z}) it results that the displacements
of the two masses under the influence of the scalar gravitational
wave are

\begin{equation}
\delta z_{b}(t)=0\label{eq: spostamento beam-splitter}\end{equation}

and

\begin{equation}
\delta z_{m}(t-v_{G}L)=-\frac{1}{2}m^{2}L\psi(t-v_{G}L).\label{eq: spostamento mirror}\end{equation}

In this way, the relative displacement, is

\begin{equation}
\delta L(t)=\delta z_{m}(t-v_{G}L)-\delta z_{b}(t)=-\frac{1}{2}m^{2}L\psi(t-v_{G}L),\label{eq: spostamento relativo}\end{equation}

Thus it results

\begin{equation}
\frac{\delta L(t)}{L}=\frac{\delta T(t)}{T}=-\frac{1}{2}m^{2}\psi(t-v_{G}L).\label{eq: strain scalare}\end{equation}

But there is the problem that, for a large separation between the
test masses (in the case of Virgo or LIGO the distance between the
beam-splitter and the mirror is three or four kilometers), the definition
(\ref{eq: spostamento relativo}) for relative displacement becomes
unphysical because the two test masses are taken at the same time
and therefore cannot be in a casual connection \cite{key-6,key-11}.
The correct definitions for our bouncing photon can be written like

\begin{equation}
\delta L_{1}(t)=\delta z_{m}(t-v_{G}L)-\delta z_{b}(t-T_{1})\label{eq: corretto spostamento B.S. e M.}\end{equation}

and

\begin{equation}
\delta L_{2}(t)=\delta z_{m}(t-v_{G}L-T_{2})-\delta z_{b}(t),\label{eq: corretto spostamento B.S. e M. 2}\end{equation}
where $T_{1}$ and $T_{2}$ are the photon propagation times for the
forward and return trip correspondingly. According to the new definitions,
the displacement of one test mass is compared with the displacement
of the other at a later time to allow for finite delay from the light
propagation. Note that the propagation times $T_{1}$ and $T_{2}$
in eqs. (\ref{eq: corretto spostamento B.S. e M.}) and (\ref{eq: corretto spostamento B.S. e M. 2})
can be replaced with the nominal value $T$ because the test mass
displacements are alredy first order in $\Phi$. Thus, for the total
change in the distance between the beam splitter and the mirror in
one round-trip of the photon, it is

\begin{equation}
\delta L_{r.t.}(t)=\delta L_{1}(t-T)+\delta L_{2}(t)=2\delta z_{m}(t-v_{G}L-T)-\delta z_{b}(t)-\delta z_{b}(t-2T),\label{eq: variazione distanza propria}\end{equation}

and in terms of the amplitude and mass of the SGW:

\begin{equation}
\delta L_{r.t.}(t)=-m^{2}L\psi(t-v_{G}L-T).\label{eq: variazione distanza propria 2}\end{equation}

The change in distance (\ref{eq: variazione distanza propria 2})
leads to changes in the round-trip time for photons propagating between
the beam-splitter and the mirror:

\begin{equation}
\frac{\delta_{1}T(t)}{T}=-m^{2}\psi(t-v_{G}L-T).\label{eq: variazione tempo proprio 1}\end{equation}

In the last calculation (variations in the photon round-trip time
which come from the motion of the test masses inducted by the scalar
gravitational wave), it was implicitly assumed that the propagation
of the photon between the beam-splitter and the mirror of our interferometer
is uniform as if it were moving in a flat space-time. But the presence
of the tidal forces indicates that the space-time is curved. As a
result another effect after the previous has to be considered, which
requires spacial separation \cite{key-6,key-11}.

For this effect we consider the interval for photons propagating along
the $z$-axis

\begin{equation}
ds^{2}=g_{00}dt^{2}+dz^{2}.\label{eq: metrica osservatore locale}\end{equation}

The condition for a null trajectory ($ds=0$) gives the coordinate
velocity of the photons

\begin{equation}
v^{2}\equiv(\frac{dz}{dt})^{2}=1+2V(t,z),\label{eq: velocita' fotone in gauge locale}\end{equation}

which to first order in $\Phi$ is approximated by

\begin{equation}
v\approx\pm[1+V(t,z)],\label{eq: velocita fotone in gauge locale 2}\end{equation}

with $+$ and $-$ for the forward and return trip respectively. Knowing
the coordinate velocity of the photon, the propagation time for its
travelling between the beam-splitter and the mirror can be defined:

\begin{equation}
T_{1}(t)=\int_{z_{b}(t-T_{1})}^{z_{m}(t)}\frac{dz}{v}\label{eq:  tempo di propagazione andata gauge locale}\end{equation}

and

\begin{equation}
T_{2}(t)=\int_{z_{m}(t-T_{2})}^{z_{b}(t)}\frac{(-dz)}{v}.\label{eq:  tempo di propagazione ritorno gauge locale}\end{equation}

The calculations of these integrals would be complicated because the
boundary $z_{m}(t)$ is changing with time. In fact it is

\begin{equation}
z_{b}(t)=\delta z_{b}(t)=0\label{eq: variazione b.s. in gauge locale}\end{equation}

but

\begin{equation}
z_{m}(t)=L+\delta z_{m}(t).\label{eq: variazione specchio nin gauge locale}\end{equation}

But, to first order in $\Phi$, this contribution can be approximated
by $\delta L_{2}(t)$ (see eq. (\ref{eq: corretto spostamento B.S. e M. 2})).
Thus, the combined effect of the varying boundary is given by $\delta_{1}T(t)$
in eq. (\ref{eq: variazione tempo proprio 1}). Then only the times
for photon propagation between the fixed boundaries $0$ and $L$
have to be calculated. Such propagation times will be denoted with
$\Delta T_{1,2}$ to distinguish from $T_{1,2}$. In the forward trip,
the propagation time between the fixed limits is

\begin{equation}
\Delta T_{1}(t)=\int_{0}^{L}\frac{dz}{v(t',z)}\approx T-\int_{0}^{L}V(t',z)dz,\label{eq:  tempo di propagazione andata  in gauge locale}\end{equation}

where $t'$ is the retardation time which corresponds to the unperturbed
photon trajectory: 

\begin{center}$t'=t-(L-z)$\end{center}

(i.e. $t$ is the time at which the photon arrives in the position
$L$, so $L-z=t-t'$).

Similiary, the propagation time in the return trip is

\begin{equation}
\Delta T_{2}(t)=T-\int_{L}^{0}V(t',z)dz,\label{eq:  tempo di propagazione andata  in gauge locale}\end{equation}

where now the retardation time is given by

\begin{center}$t'=t-z$.\end{center}

The sum of $\Delta T_{1}(t-T)$ and $\Delta T_{2}(t)$ gives the round-trip
time for photons traveling between the fixed boundaries. Then the
deviation of this round-trip time (distance) from its unperturbed
value $2T$ is

\begin{equation}
\delta_{2}T(t)=\int_{0}^{L}[V(t-2T+z,z)+V(t-z,z)]dz.\label{eq: variazione tempo proprio lungo z 2}\end{equation}

From eqs. (\ref{eq:potenziale in gauge Lorentziana generalizzato})
and (\ref{eq: variazione tempo proprio lungo z 2}) it results:

\begin{equation}
\begin{array}{c}
\delta_{2}T(t)=\frac{1}{2}m^{2}\int_{0}^{L}[\int_{0}^{z}\Phi(t-2T+a-v_{G}a)ada+\int_{0}^{z}\Phi(t-a-v_{G}a)ada]dz=\\
\\=\frac{1}{4}m^{2}\int_{0}^{L}[\Phi(t-v_{G}z-2T+z)+\Phi(t-v_{G}z-z)]z^{2}dz+\\
\\-\frac{1}{4}m^{2}\int_{0}^{L}[\int_{0}^{z}\Phi'(t-2T+a-v_{G}a)z^{2}da+\int_{0}^{z}\Phi'(t-a-v_{G}a)z^{2}da]dz,\end{array}\label{eq: variazione tempo proprio lungo z 2.2}\end{equation}

Thus the total round-trip proper distance in presence of the scalar
gravitational wave is:

\begin{equation}
T=2T+\delta_{1}T+\delta_{2}T.\label{eq: round-trip  totale in gauge locale}\end{equation}

Now, to obtain the interferometer response function of the massive
scalar field, the analysis can be transled in the frequency domine.

Using the Fourier transform of $\psi$ defined from 

\begin{equation}
\tilde{\psi}(\omega)=\int_{-\infty}^{\infty}dt\psi(t)\exp(i\omega t),\label{eq: trasformata di fourier2}\end{equation}
 eq. (\ref{eq: variazione tempo proprio 1}) can be rewritten like:

\begin{equation}
\frac{\delta_{1}\tilde{T}(\omega)}{T}=-m^{2}\Upsilon_{1}^{*}(\omega)\tilde{\psi}(\omega)\label{eq: fourier 1 lungo z}\end{equation}

with 

\begin{equation}
\Upsilon_{1}^{*}(\omega)=\exp[i\omega(1+v_{G})L].\label{eq: risposta 1 lungo z}\end{equation}
But, from a theorem about Fourier transforms, it's simple to obtain:

\begin{equation}
\tilde{\psi}(\omega)=-\frac{\tilde{\Phi}(\omega)}{\omega^{2}},\label{eq: Teorema di Fourier}\end{equation}

where

\begin{equation}
\tilde{\Phi}(\omega)=\int_{-\infty}^{\infty}dt\Phi(t)\exp(i\omega t).\label{eq: trasformata di fourier}\end{equation}

is the Fourier transform of our scalar field.

Then it results:

\begin{equation}
\frac{\delta_{1}\tilde{T}(\omega)}{T}=\frac{m^{2}}{\omega^{2}}\Upsilon_{1}^{*}(\omega)\tilde{\Phi}(\omega),\label{eq: delta t su t finale}\end{equation}
and, defining:

\begin{equation}
\Upsilon_{1}\equiv\frac{m^{2}}{\omega^{2}}\Upsilon_{1}^{*}(\omega)=(1-v_{G}^{2})\Upsilon_{1}^{*}(\omega),\label{eq: def.  gamma1}\end{equation}

we obtain:

\begin{equation}
\frac{\delta_{1}\tilde{T}(\omega)}{T}=\Upsilon_{1}(\omega)\tilde{\Phi}(\omega).\label{eq: delta t su t finale 2}\end{equation}

On the other hand eq. (\ref{eq: variazione tempo proprio lungo z 2.2})
can be rewritten in the frequency space like:

\begin{equation}
+\begin{array}{c}
\delta_{2}\tilde{T}(\omega)=\frac{1}{2\omega(v_{G}^{2}-1)^{2}}[\exp[2i\omega L](v_{G}+1)^{3}(-2i+\omega L(v_{G}-1)+\\
\\2\exp[i\omega L(1+v_{G})](6iv_{G}+2iv_{G}^{3}-\omega L+\omega Lv_{G}^{4})+\\
\\+(v_{G}+1)^{3}(-2i+\omega L(v_{G}+1))]\tilde{\Phi}(\omega).\end{array}\label{eq: variazione tempo proprio lungo z 2.3}\end{equation}

Now 

\begin{equation}
\frac{\delta_{2}\tilde{T}(\omega)}{T}=\Upsilon_{2}(\omega)\tilde{\Phi}(\omega),\label{eq: fourier 2 lungo z}\end{equation}

can be put, with 

\begin{equation}
\begin{array}{c}
\Upsilon_{2}(\omega)=\frac{1}{2\omega L(v_{G}^{2}-1)^{2}}[\exp[2i\omega L](v_{G}+1)^{3}(-2i+\omega L(v_{G}-1)+\\
\\2\exp[i\omega L(1+v_{G})](6iv_{G}+2iv_{G}^{3}-\omega L+\omega Lv_{G}^{4})+\\
\\+(v_{G}+1)^{3}(-2i+\omega L(v_{G}+1))].\end{array}\label{eq: risposta 2 lungo z}\end{equation}
Because it is

\begin{equation}
\Upsilon_{l}(\omega)=\Upsilon_{1}(\omega)+\Upsilon_{2}(\omega),\label{eq: risposta totale lungo z}\end{equation}
from eqs. (\ref{eq: risposta 1 lungo z}), (\ref{eq: def.  gamma1})
and (\ref{eq: risposta 2 lungo z}) it results that the function 

\begin{equation}
\begin{array}{c}
\Upsilon_{l}(\omega)=(1-v_{G}^{2})\exp[i\omega L(1+v_{G})]+\frac{1}{2\omega L(v_{G}^{2}-1)^{2}}[\exp[2i\omega L](v_{G}+1)^{3}(-2i+\omega L(v_{G}-1)+\\
\\2\exp[i\omega L(1+v_{G})](6iv_{G}+2iv_{G}^{3}-\omega L+\omega Lv_{G}^{4})+\\
\\+(v_{G}+1)^{3}(-2i+\omega L(v_{G}+1))].\end{array}\label{eq: risposta totale lungo z due}\end{equation}

is the response function of an arm of our interferometer located in
the $z$-axis, due to the longitudinal component of the massive scalar
gravitational wave propagating in the same direction of the axis.

For $v_{G}\rightarrow1$ it is $\Upsilon_{l}(\omega)\rightarrow0$. 

The longitudinal response function (\ref{eq: risposta totale lungo z due})
has been obtained in function of the group velocity of the wave-packet.
But, putting eq. (\ref{eq: velocita' di gruppo 2}) into eq. (\ref{eq: risposta totale lungo z due})
it also results \begin{equation}
\begin{array}{c}
\Upsilon_{l}(\omega)=\frac{1}{m^{4}\omega^{2}L}(\frac{1}{2}(1+\exp[2i\omega L])m^{2}\omega^{2}L(m^{2}-2\omega^{2})+\\
\\-i\exp[2i\omega L]\omega^{2}\sqrt{-m^{2}+\omega^{2}}(4\omega^{2}+m^{2}(-1-iL\omega))+\\
\\+\omega^{2}\sqrt{-m^{2}+\omega^{2}}(-4i\omega^{2}+m^{2}(i+\omega L))+\\
\\+\exp[iL(\omega+\sqrt{-m^{2}+\omega^{2}})](m^{6}L+m^{4}\omega^{2}L+8i\omega^{4}\sqrt{-m^{2}+\omega^{2}}+\\
\\+m^{2}(-2L\omega^{4}-2i\omega^{2}\sqrt{-m^{2}+\omega^{2}}))+2\exp[i\omega L]\omega^{3}(-3m^{2}+4\omega^{2})\sin[\omega L]).\end{array}\label{eq: risposta totale lungo z massa}\end{equation}

Using the second of equations (\ref{eq: definizione}) equation (\ref{eq: risposta totale lungo z massa})
reads \begin{equation}
\begin{array}{c}
\Upsilon_{l}(\omega)=\frac{1}{(\frac{\beta R_{0}}{2R_{0}-12})^{4}\omega^{2}L}(\frac{1}{2}(1+\exp[2i\omega L])\frac{\beta R_{0}}{2R_{0}-12}\omega^{2}L(\frac{\beta R_{0}}{2R_{0}-12}-2\omega^{2})+\\
\\-i\exp[2i\omega L]\omega^{2}\sqrt{-\frac{\beta R_{0}}{2R_{0}-12}+\omega^{2}}(4\omega^{2}+\frac{\beta R_{0}}{2R_{0}-12}(-1-iL\omega))+\\
\\+\omega^{2}\sqrt{-\frac{\beta R_{0}}{2R_{0}-12}+\omega^{2}}(-4i\omega^{2}+\frac{\beta R_{0}}{2R_{0}-12}(i+\omega L))+\\
\\+\exp[iL(\omega+\sqrt{-\frac{\beta R_{0}}{2R_{0}-12}+\omega^{2}})]((\frac{\beta R_{0}}{2R_{0}-12})^{3}L+(\frac{\beta R_{0}}{2R_{0}-12})^{4}\omega^{2}L+8i\omega^{4}\sqrt{-\frac{\beta R_{0}}{2R_{0}-12}+\omega^{2}}+\\
\\+\frac{\beta R_{0}}{2R_{0}-12}(-2L\omega^{4}-2i\omega^{2}\sqrt{-\frac{\beta R_{0}}{2R_{0}-12}+\omega^{2}}))+2\exp[i\omega L]\omega^{3}(-3\frac{\beta R_{0}}{2R_{0}-12}+4\omega^{2})\sin[\omega L]).\end{array}\label{eq: risposta totale lungo z R}\end{equation}

In this way it appears that, the detection of this longitudinal component
of the general gravitational waves (\ref{eq: perturbazione totale})
arising from the action (\ref{eq: high order 1}) at various frequencies
can be connected, in principle, to the Ricci scalar.

In figures 4, 5 and 6 are shown the response functions (\ref{eq: risposta totale lungo z due})
for an arm of the Virgo interferometer ($L=3Km$) for $v_{G}=0.1$
(non-relativistic case), $v_{G}=0.9$ (relativistic case) and $v_{G}=0.999$
(ultra-relativistic case). We see that in the non-relativistic case
the signal is stronger as it could be expected (for $m\rightarrow0$
we expect$\Upsilon_{l}(\omega)\rightarrow0$). In figures 7, 8, and
9 the same response functions are shown for the Ligo interferometer
($L=4Km$).

\begin{figure}
\includegraphics{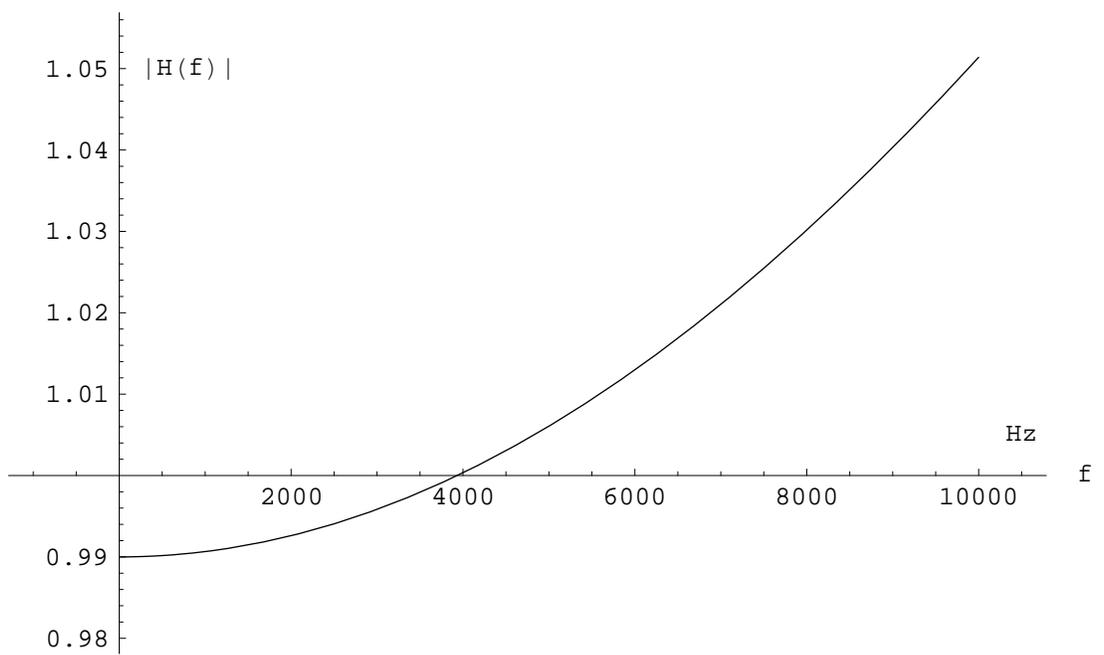}

\caption{the absolute value of the longitudinal response function (\ref{eq: risposta totale lungo z due})
of the Virgo interferometer ($L=3Km$) to a SGW propagating with a
speed of $0.1c$ (non relativistic case). }
\end{figure}
\begin{figure}
\includegraphics{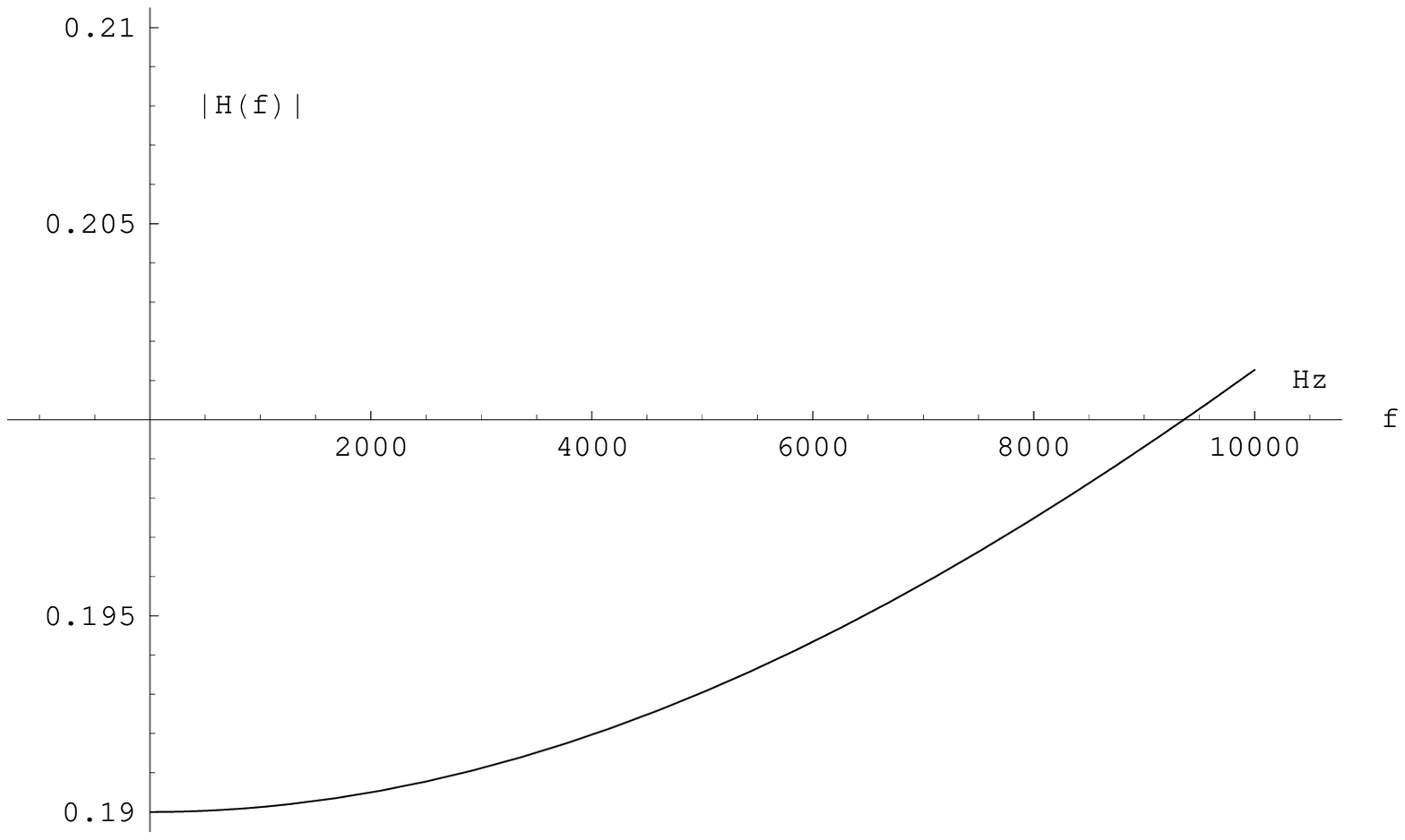}

\caption{the absolute value of the longitudinal response function (\ref{eq: risposta totale lungo z due})
of the Virgo interferometer ($L=3Km$) to a SGW propagating with a
speed of $0.9$ (relativistic case). }
\end{figure}
\begin{figure}
\includegraphics{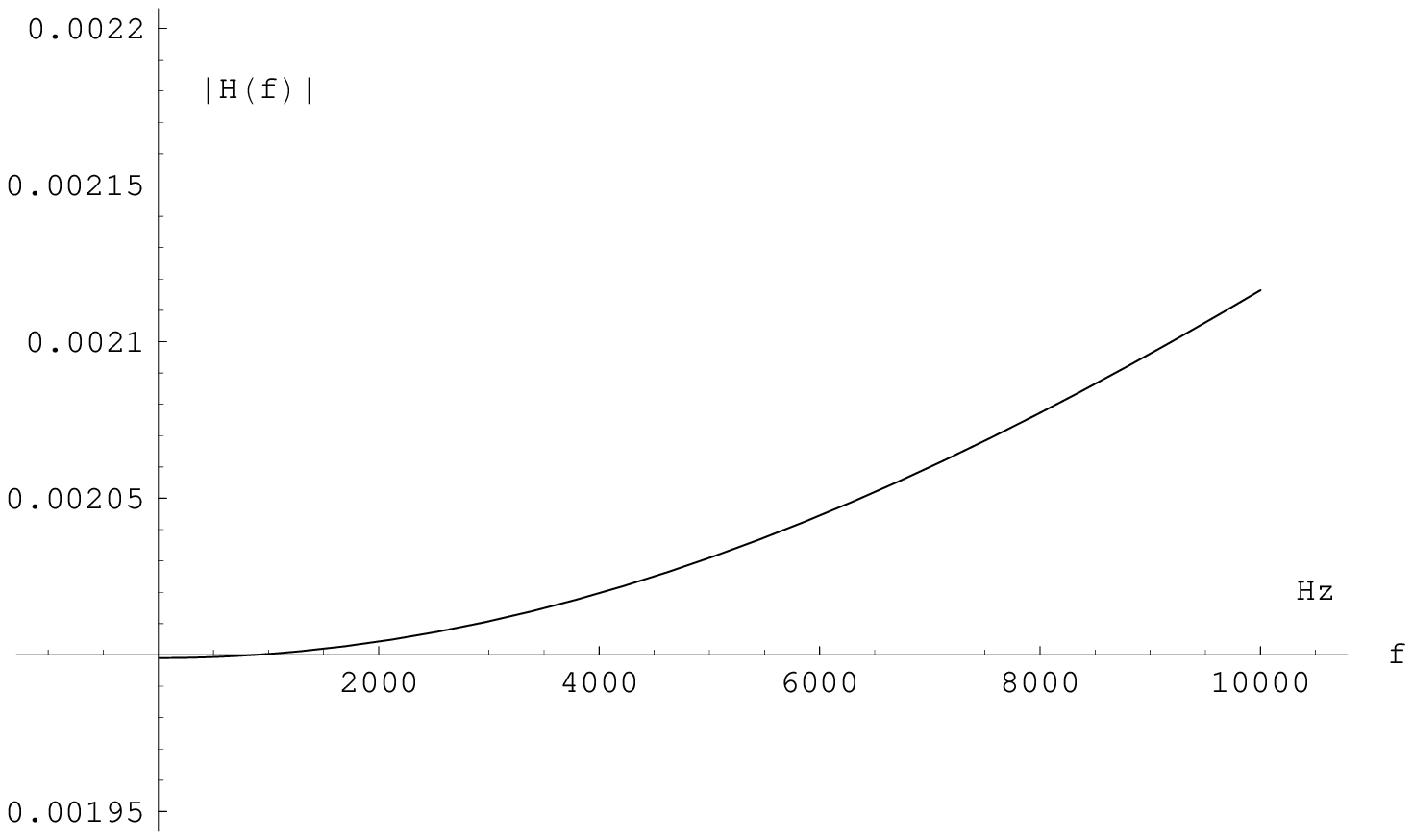}

\caption{the absolute value of the longitudinal response function (\ref{eq: risposta totale lungo z due})
of the Virgo interferometer ($L=3Km$) to a SGW propagating with a
speed of $0.999$ (ultra relativistic case). }
\end{figure}
\begin{figure}
\includegraphics{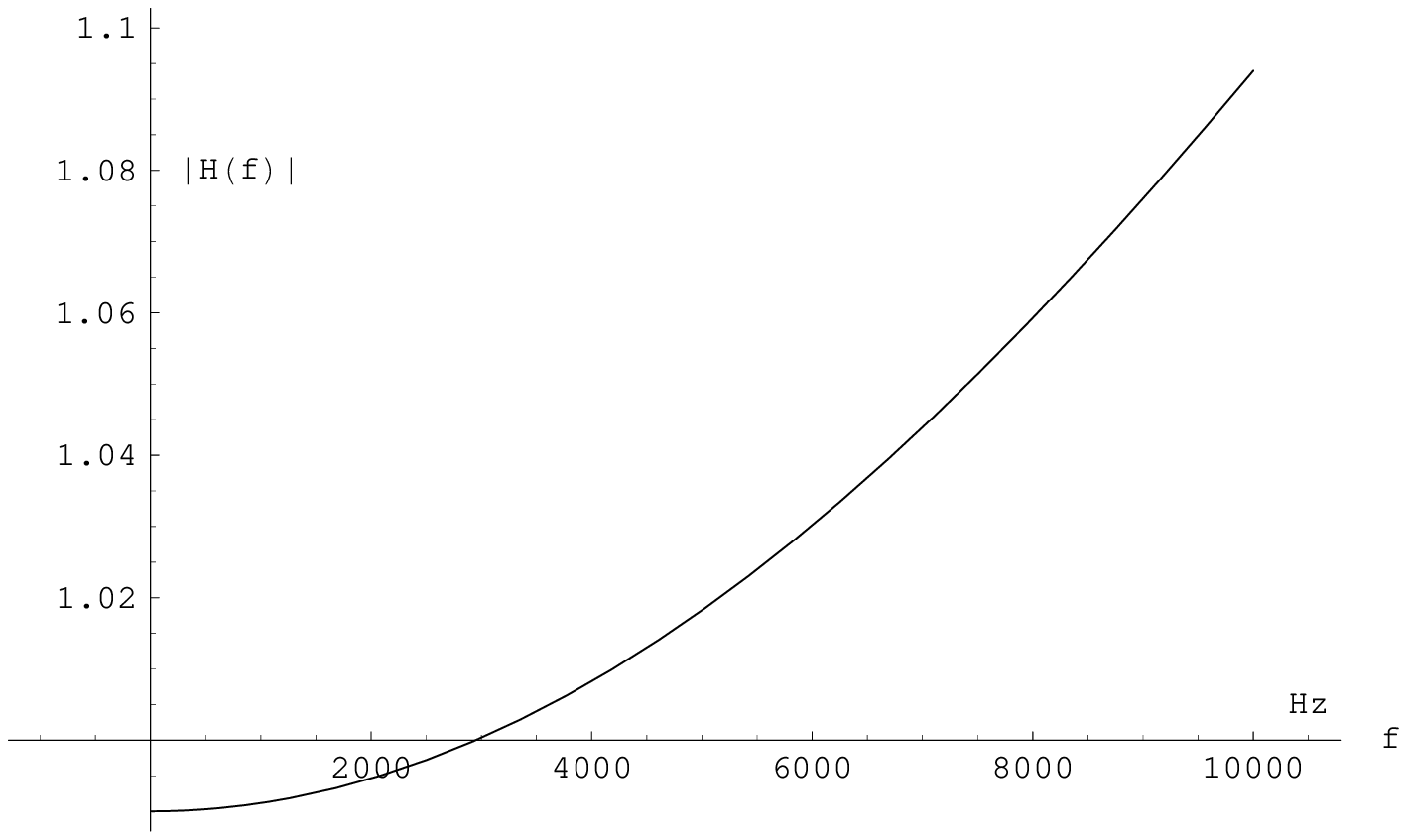}

\caption{the absolute value of the longitudinal response function (\ref{eq: risposta totale lungo z due})
of the LIGO interferometer ($L=4Km$) to a SGW propagating with a
speed of $0.1c$ (non relativistic case). }
\end{figure}
\begin{figure}
\includegraphics{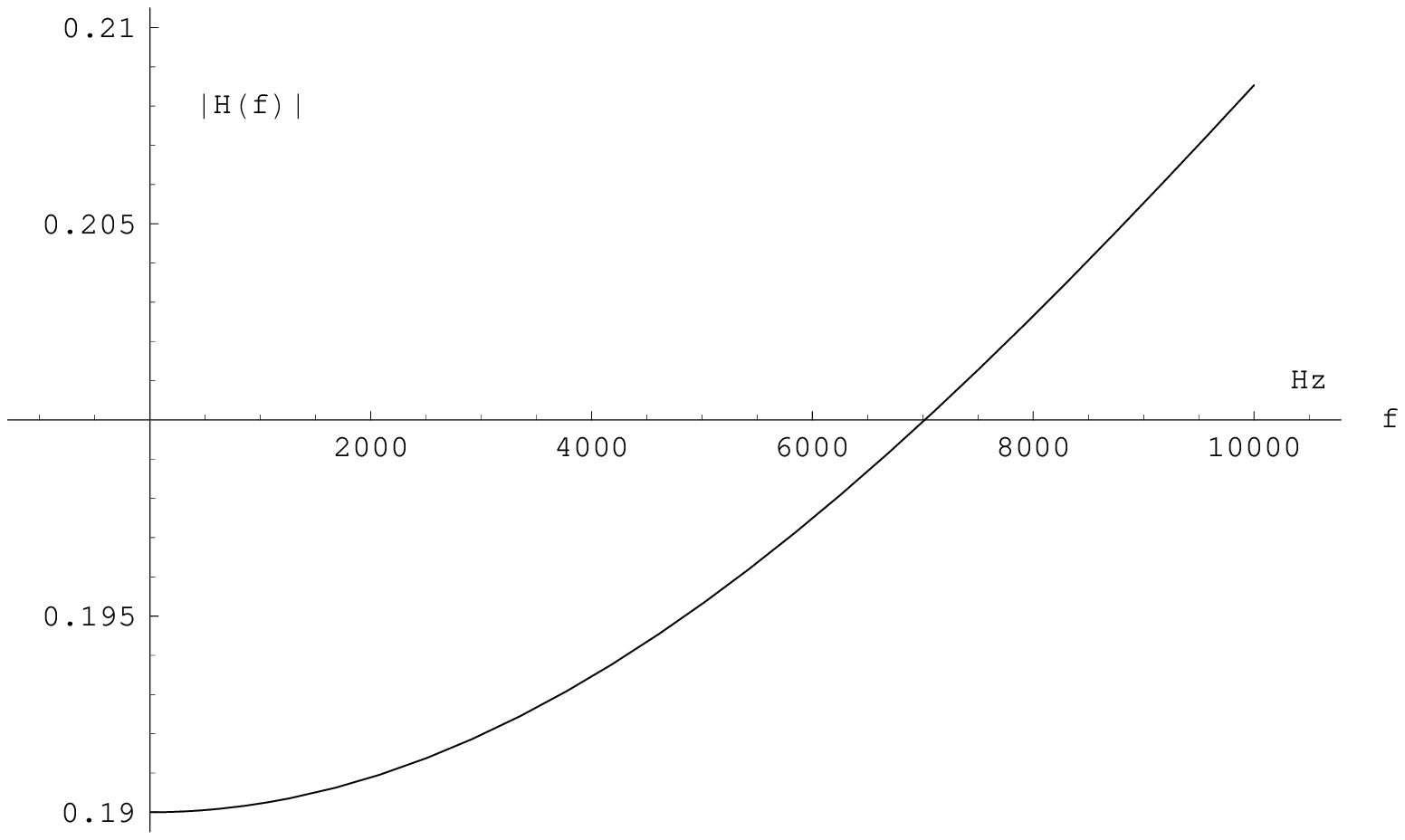}

\caption{the absolute value of the longitudinal response function (\ref{eq: risposta totale lungo z due})
of the LIGO interferometer ($L=4Km$) to a SGW propagating with a
speed of $0.9c$ (relativistic case). }
\end{figure}
\begin{figure}
\includegraphics{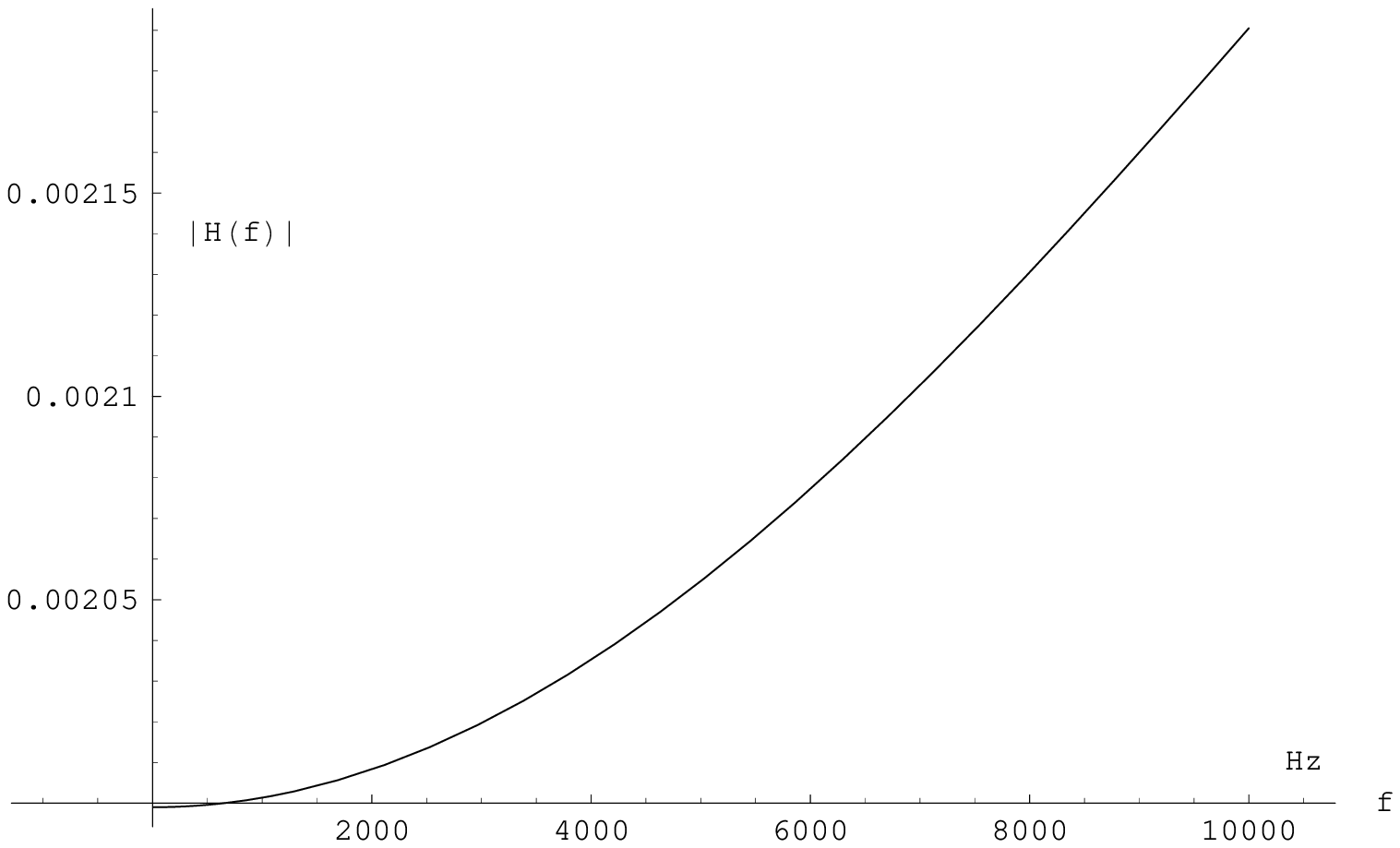}

\caption{the absolute value of the longitudinal response function (\ref{eq: risposta totale lungo z due})
of the LIGO interferometer ($L=4Km$) to a SGW propagating with a
speed of $0.999c$ (ultra relativistic case). }
\end{figure}

\section{Conclusions}

The production and the potential detection with interferometers of
a hypotetical massive \textit{scalar} component of gravitational radiation
which arises from the $R^{-1}$ theory of gravity has been shown.
This agrees with the formal equivalence between high order theories
of gravity and scalar tensor gravity that is well known in literature.

First it has been shown that a massive scalar mode of gravitational
radiation arises from the particular action of he $R^{-1}$ theory
of gravity. 

After this, it has been shown that the fact that gravitational waves
can have a massive component generates a longitudinal force in addition
of the transverse one which is proper of the massless case. 

Then, the potential interferometric detection of this longitudinal
component has been analyzed and the response of an interferometer
has been computed. It has also been shown that this longitudinal response
function is directly connected with the Ricci curvature scalar. 

In the analysis of the response of the interferometers an analysis
parallel to the one seen in \cite{key-6} and \cite{key-11} has been
used.

As a final remark we emphasize that, an investigation on scalar components
of gravitational waves opens to the possibility of using the signals
seen from interferometers to understand which is the correct theory
of gravitation, while the presence of the mass could also have important
applications in cosmology because the fact that gravitational waves
can have mass could give a contribution to the dark matter of the
Universe.

\section{Acknowledgements}

I would like to thank Salvatore Capozziello, Maria Felicia De Laurentis
and Franceso Rubanu for helpful advices during our work. I have to
thank the European Gravitational Observatory (EGO) consortium for
the using of computing facilities.

\end{document}